\documentclass[11pt,a4paper]{article}
\usepackage{jcappub}

\newcommand{\bea}{\begin{eqnarray}}
\newcommand{\eea}{\end{eqnarray}}
\newcommand{\beq}{\begin{eqnarray}}
\newcommand{\eeq}{\end{eqnarray}}
\newcommand{\pd}{\partial}
\newcommand{\nd}{\nabla}
\newcommand{\ba}{\begin{eqnarray}}
\newcommand{\ea}{\end{eqnarray}}
\newcommand{\nn}{\nonumber\\}
\newcommand{\be}{\begin{equation}}
\newcommand{\ee}{\end{equation}}
\newcommand{\mc}{\mathcal}
\newcommand{\tx}{\mathrm}

\newcommand{\tr}{\tilde{r}}

\title{Asymptotically flat black holes in Horndeski theory and beyond}

\author[a]{E.~Babichev,} 
\author[a]{C.~Charmousis,}
\author[a]{A.~Leh\'ebel} 

\affiliation[a]{Laboratoire de Physique Th\'eorique, CNRS, Univ. Paris-Sud, \\ Universit\'e Paris-Saclay, 91405 Orsay, France}
\emailAdd{eugeny.babichev@th.u-psud.fr}
\emailAdd{christos.charmousis@th.u-psud.fr}
\emailAdd{antoine.lehebel@th.u-psud.fr}

\abstract{We find spherically symmetric and static black holes in shift-symmetric Horndeski and beyond Horndeski theories. They are asymptotically flat and sourced by a non trivial static scalar field. The first class of solutions is constructed in such a way that the Noether current associated with shift symmetry vanishes, while the scalar field cannot be trivial. This in certain cases leads to hairy black hole solutions (for the quartic Horndeski Lagrangian), and in others to singular solutions (for a Gauss-Bonnet term).  Additionally, we find the general spherically symmetric and static solutions for a pure quartic Lagrangian, the metric of which is Schwarzschild. We show that under two requirements on the theory in question, any vacuum GR solution is also solution to the quartic theory. As an example, we show that a Kerr black hole with a non-trivial scalar field is an exact solution to these theories.}

\begin{document}

\maketitle


\section{Introduction}

General Relativity (GR) accounts for gravitational phenomena for local distance scales of the order of up to 30 or so astronomical units. These range from weak tests in the solar system, to stronger tests as those of binary pulsars \cite{Will:2014kxa}, and very recently of coalescing black holes exhibiting each around 30 solar masses \cite{Abbott:2016blz}. Nevertheless, at cosmological distance scales, of the order of $10^{15}$ greater than the local ones, one has to introduce an unexplained dark energy component in order to account for the present acceleration of the expansion of the universe. This is still not enough to relieve the tension involving GR and missing matter. Even at galactic scales, a component of dark matter has to be added, of completely different nature to dark energy. The origin of dark matter, although far less exotic than dark energy, still evades detection in the LHC particle colliders, as well as numerous astroparticle detectors. This tension in between undetected sources of energy and GR assumes that at such astrophysical or cosmological scales, gravity strictly obeys Einstein's field equations. In fact, only a mere $4\%$ of the present matter content in our Universe is some detected and understood matter source. Dark energy is usually assumed to be driven by a minute cosmological constant which also lacks theoretical explanation on why it is fine tuned to such a tiny value. To make matters worse, the whole discussion on dark energy is plagued by the big cosmological constant problem: theoretically, the cosmological constant should be huge and unstable to radiative corrections. One therefore is quite naturally led to question the validity of GR at very large distance scales, deep in the infrared. Additionally, modified gravity theories provide some interesting mechanisms to bifurcate the cosmological constant problem, at least at the classical level (see for example \cite{Charmousis:2011bf, Charmousis:2011ea, Appleby:2012rx, Babichev:2016kdt})\footnote{The crucial field theoretic question of radiative stability of the cosmological constant has been recently addressed in a series of interesting papers \cite{Kaloper:2013zca,Kaloper:2014dqa}.}. 

Among modified gravity theories, a robust alternative to GR is offered by scalar-tensor theories which possess an additional scalar degree of freedom. Scalar-tensor theories include key characteristics of modified gravity as most gravity theories have some scalar-tensor limit. The most general scalar-tensor theory with second derivative equations of motion was found by Horndeski back in the $70$'s. Recently, in a sequence of papers, Gleyzes \textit{et al.} realized that Horndeski theory was not the only physically relevant alternative \cite{Gleyzes:2014dya,Gleyzes:2014qga} for scalar-tensor theories. Although higher derivative equations of motion generically lead to unstable theories, it can so happen that certain theories have higher derivatives and still be physically sensible. The solution to this puzzling property is that there exist theories with higher derivative equations of motion which are degenerate, and still have the canonical number of propagating degrees of freedom, thus evading ghosts~\cite{Gleyzes:2014dya,Gleyzes:2014qga,Zumalacarregui:2013pma,Deffayet:2015qwa,Crisostomi:2016tcp,Langlois:2015skt,BenAchour:2016fzp}.  These theories are called beyond Horndeski theories. We will consider black hole solutions of Horndeski and beyond Horndeski theories in this paper. With the advent of gravitational wave detection and the upcoming telescope array experiments, such as the Event Horizon Telescope \cite{Ricarte:2014nca}, compact gravitating objects are of great importance for they may provide smoking gun signals of modified gravity theories. This will be especially true if there exist solutions that evade no-hair theorems (for a recent review, see \cite{Herdeiro:2015waa}) with genuine gravitating scalars modifying the GR predictions\footnote{In scalar-tensor theories, usual GR black holes with a trivial scalar field are also solutions to the field equations of the scalar-tensor theory although a priori they are phenomenologically less interesting.}. 
%

It has been shown recently that Horndeski and beyond Horndeski theories allow for black hole solutions with static non-Schwarzschild metric and non-trivial scalar field. 
Both these modifications with respect to GR black holes open an interesting possibility to test the theory with upcoming observations. 
One approach to construct black hole solutions, entertained in~\cite{Babichev:2013cya}, was to allow for a radially and time-dependent scalar field.
This assumption leads to a family of black hole solutions with static metric and a non-trivial time-dependent scalar field profile~\cite{Babichev:2013cya,Babichev:2015qma,Charmousis:2015aya,Kobayashi:2014eva,Babichev:2016kdt,Babichev:2016fbg,Charmousis:2014zaa}. 
On the other hand, if the scalar field is assumed to be static, a no-hair theorem dictates that, generically, static and asymptotically flat black hole solutions of shift-symmetric Horndeski theories must have GR form and a constant scalar field~\cite{Hui:2012qt}. 
The no-hair theorem contains, as usual, several assumptions. If one or more assumptions is broken, some examples of hairy black holes can be found within Horndeski theory~\cite{Rinaldi:2012vy,Anabalon:2013oea,Minamitsuji:2013ura,Sotiriou:2014pfa,Maselli:2015yva}.

In this paper, we examine in detail shift-symmetric Horndeski and beyond Horndeski theories which allow for static and asymptotically flat black holes with a static scalar field.
Our motivations here are the following. 
First, we would like to find out whether there exist shift-symmetric theories that allow for static and asymptotically flat black holes with a regular Noether current. 
Indeed, in spite of previous attempts, no such consistent solution has been exhibited up to now. 
On the technical side, we shall check if the hypotheses of the no-hair theorem are indeed minimal: 
we would like to provide an example of a hairy black hole whenever one single hypothesis of the no-hair theorem is broken. 
Finally, we also provide examples of black hole solutions that can be used for benchmark tests of GR.

We start with refining the no-hair theorem for Horndeski theory~\cite{Hui:2012qt}, including the beyond Horndeski terms as well. 
Then we proceed in two different ways to select a suitable theory, each time breaking only minimally the assumptions of the no-hair theorem.
In the first category, we fix six particular Lagrangians (four in the case of pure Horndeski theory) which may potentially support static black hole solutions. 
As an illustrative example, we present a new black hole solution for a quartic Galileon theory, with everywhere finite norm of the scalar current.
Another interesting Lagrangian in this class is the Galileon linearly coupled to the Gauss-Bonnet term, which has been discussed before in~\cite{Sotiriou:2014pfa,Sotiriou:2013qea}.
For this theory, however, we do not find black holes with a finite norm of the scalar current. 
A different approach is used to find a second category of theories supporting static and asymptotically flat black hole solutions. 
We aim to find here stealth solutions, i.e. those having GR metric solutions, but nevertheless a nontrivial scalar field configuration. 
In this category we find a family of quartic Lagrangians which must satisfy some precise conditions. 

In the next section, we start with the action and equations on motion for beyond Horndeski theory, 
and then assume a static and spherically symmetric ansatz.  
We then revise the no-hair theorem to include the beyond Horndeski Lagrangians, and 
select Lagrangians which a priori may give asymptotically flat black hole solutions. 
Particular examples of the theories are investigated in the following sections.
In Sec.~\ref{sec:nanalytic}, we find and examine in detail some static solution for a model that includes a standard kinetic term and a specific quartic Lagrangian. We also examine the consequences of allowing the scalar field to become time-dependent for this specific model. Section~\ref{sec:GB} discusses the model with a linear coupling between the Gauss-Bonnet invariant and the scalar field. Section~\ref{sec:quartic} is devoted to the generic solution for a purely quartic beyond Horndeski action. Finally, Sec.~\ref{sec:conc} contains our conclusions.

\section{Theories potentially containing black holes with hair}
\label{sec:escape}

We will focus on the subclass of (beyond) Horndeski theory which possesses shift symmetry $\phi \rightarrow \phi + constant$. It involves six arbitrary functions of the kinetic term $X=- \pd^\mu \phi \pd_\mu \phi/2$. They are denoted as $G_2, G_3, G_4, G_5$ for ordinary Horndeski, as well as $F_4$ and $F_5$ for beyond, and combine in the action as follows:
\bea
\label{action}
S &=& \int\, \sqrt{-g}\, d^4x \left(\mc{L}_2+\mc{L}_3+\mc{L}_4+\mc{L}_5 +\mc{L}^{\rm bH}_4+\mc{L}^{\rm bH}_5 \right),
\eea
with
\bea
\label{ssgg}
\mc{L}_2 &=& G_2(X) ,
\\
\mc{L}_3 &=& -G_3(X) \Box \phi ,
\\
\mc{L}_4 &=& G_4(X) R + G_{4X} \left[ (\Box \phi)^2 -(\nabla_\mu\nabla_\nu\phi)^2 \right] ,
\\
\mc{L}_5 &=& G_5(X) G_{\mu\nu}\nabla^\mu \nabla^\nu \phi - \frac{1}{6} G_{5X} \big[ (\Box \phi)^3 - 3\Box \phi(\nabla_\mu\nabla_\nu\phi)^2 \nn
&~&~~ + 2(\nabla_\mu\nabla_\nu\phi)^3 \big],
\\
\mc{L}_4^{\tx{bH}} &=& F_4(X) \epsilon^{\mu \nu \rho \sigma} {\epsilon^{\alpha \beta \gamma}}_\sigma \nd_\mu\phi \nd_\alpha\phi \nd_\nu \nd_\beta\phi \nd_\rho \nd_\gamma\phi ,
\\
\mc{L}_5^{\tx{bH}} &=& F_5(X) \epsilon^{\mu \nu \rho \sigma} \epsilon^{\alpha \beta \gamma \delta} \nd_\mu\phi \nd_\alpha\phi \nd_\nu \nd_\beta\phi \nd_\rho \nd_\gamma\phi \nd_\sigma \nd_\delta\phi ,
\eea
where a subscript $X$ stands for the derivative with respect to $X$, $R$ is the Ricci scalar, $G_{\mu\nu}$ is the Einstein tensor, $(\nabla_\mu\nabla_\nu\phi)^2 = \nabla_\mu\nabla_\nu\phi \nabla^\nu\nabla^\mu\phi$ and $(\nabla_\mu\nabla_\nu\phi)^3=\nabla_\mu\nabla_\nu\phi \nabla^\nu\nabla^\rho\phi \nabla_\rho\nabla^\mu\phi$. Shift symmetry is accompanied by a Noether current:
\begin{eqnarray}
\label{covcur}
J^\mu&\equiv&\frac{1}{\sqrt{-g}}\,
\frac{\delta S[\phi]}{\delta (\partial_\mu\phi)}.
\end{eqnarray}
We will suppose that both spacetime and  scalar are spherically symmetric and static:
\bea
\tx{d} s^2 &=& -h(r) \tx{d} t^2 + \dfrac{\tx{d} r^2}{f(r)} + r^2 (\tx{d}\theta^2 + \sin^2 \theta \tx{d}\varphi^2),\\
\phi &=& \phi(r).
\label{eq:ansatz}
\eea
Note that shift symmetry allows to consistently relax staticity of the scalar field (up to a linear time dependence) as in~\cite{Babichev:2013cya} and still have consistent equations of motion \cite{Babichev:2015rva}.
We will come back to this possibility, but for now we hold on to staticity. The only non-vanishing component of the current is then the radial one:
\bea
\label{eq:Jrexpr}
J^r=&-&f \phi' G_{2X} - f\dfrac{r h' + 4 h}{r h} X G_{3X} + 2 f \phi' \dfrac{f h - h + r f h'}{r^2 h}G_{4X} + 4 f^2 \phi' \dfrac{h +rh' }{r^2h} X G_{4XX} 
\nn
&-&f h' \dfrac{1 - 3 f}{r^2h} X G_{5X} +2 \dfrac{h' f^2}{r^2h} X^2 G_{5XX}+ 8 f^2 \phi' \dfrac{h +rh' }{r^2h} X (2 F_4+X F_{4X})
\nn
&-& 12 \dfrac{f^2 h'}{rh} X^2 (5 F_5+2 X F_{5X})
,\eea
where a prime stands for a derivative with respect to the radial coordinate $r$. 
In the case of Horndeski theory, i.e. when $F_i=0$, Ref.~\cite{Hui:2012qt} has shown that, under some assumptions, 
black holes have no hair (see~\cite{Sotiriou:2014pfa,Babichev:2016rlq} for more details).
These no-hair arguments can be adjusted to the case of beyond Horndeski theory.
Indeed, one of the key assumptions of the no-hair theorem, 
besides asymptotic flatness and vanishing derivative of $\phi'$ at infinity, is 
that the norm of the current $J^\mu J_\mu$ is finite down to the horizon~\cite{Hui:2012qt}. Reference~\cite{Hui:2012qt} considers this hypothesis as a physical one, as it is quite natural for matter to couple to the current as the black hole is formed. 
This immediately leads to $J^r=0$ for all radii.
In order to achieve the no-hair result, one needs to add two extra conditions on the form of the Lagrangian:
\bea
&\phantom{~}&\text{1.\:The functions $G_i$ and $F_i$ are such that their $X$-derivatives contain only positive or zero}
\nn
&\phantom{~}&\text{~~~~powers of $X$ when $X\rightarrow0$ (as $r\rightarrow + \infty$),}
\nn
&\phantom{~}&\text{2.\:There must be a canonical kinetic term $X$ in the action.}
\label{eq:hyp}
\eea
Then the field equations result in a Schwarzschild-isometric black hole and a constant scalar field. 
Relaxing assumptions of mathematical theorems usually leads to generalizing the validity of the theorem in question. In this case, the most interesting aspect of this no-hair theorem is that failing each the hypotheses leads to constructing hairy black holes! 
In this paper we will pursue this aim, breaking either of the latter two assumptions. 

As a first possibility, let us break assumption 1, i.e.  we allow $G_i$ ($i\neq 2$) and $F_i$ to be arbitrary functions of $X$, so that no assumptions are made on their derivatives. At the same time, we keep the standard kinetic term, so we do not break assumption 2. It is natural to do this at the level of spherical symmetry, keeping the regularity condition $J^r=0$. Given Eq.~(\ref{eq:Jrexpr}) and since  $G_2 \supseteq X$, the only way to evade a trivial solution $\phi'=0$ is to make one of the $G_i$ or $F_i$ pieces independent of $\phi'$\footnote{It could also occur that $J^r$ contains negative powers of $\phi'$. However, such solutions would acquire an infinite current when approaching Minkowski vacuum.}. This way, $\phi'$ (appearing in the standard kinetic term) will be forced to a non-trivial value from the condition $J^r=0$.
This occurs for particular choices of $G_i$ and $F_i$. 
A careful examination of (\ref{eq:Jrexpr}) reveals that for each function $G_i$ and $F_i$, there exists such a choice:
\bea
G_2 &\supseteq& \sqrt{-X},
\nn
G_3 &\supseteq& \tx{ln}|X|,
\nn
G_4 &\supseteq& \sqrt{-X},
\nn
G_5 &\supseteq& \tx{ln}|X|,
\nn
F_4 &\supseteq& (-X)^{-3/2},
\nn
F_5 &\supseteq& X^{-2}.
\label{eq:nanalytic}
\eea
A theory with an action involving the standard kinetic term, $G_2 \supseteq X$,  additionally to one of the above Lagrangians, has the potential to possess a non-GR static and asymptotically flat black hole solution, with regular behavior for the current. We should emphasize that it suffices for one of the above terms to have such a form (see the example in the next section). This guarantees that $\phi'$ is not trivial but does not guarantee the existence of a black hole solution. It is a necessary condition.
As specific examples, in Sec.~\ref{sec:nanalytic}, we present exact black hole solutions for the theories with $G_4 \propto \sqrt{-X}$ and $F_4 \propto (-X)^{-3/2}$ respectively. On the contrary, as we will see, the theory with $G_5 \propto \tx{ln}|X|$ does not have a black hole solution for $J^r=0$. 

The other possibility to look for hairy black hole solutions is to give up assumption~2 in the above, 
i.e. to consider a theory which only involves one or several Galileon terms, none of them being the canonical kinetic term. 
This possibility is considered in Sec.~\ref{sec:quartic}, where we find a family of Lagrangians, for which the metric is GR-like, 
but the scalar field is nontrivial. These black hole solutions are similar to the stealth solutions found in~\cite{Babichev:2013cya} 
in the context of the theory with a specific quartic Lagrangian (see also~\cite{Kobayashi:2014eva,Babichev:2015qma,Babichev:2016kdt} for generalizations to other Lagrangians). 
In contrast to the solutions found in these papers, however, the solutions we present in~Sec.~\ref{sec:quartic} have time-independent scalar field configurations.

\section{Black hole solutions in Horndeski and beyond quartic models}
\label{sec:nanalytic}

In this section, we present two black hole solutions: one in the class of Horndeski theory and one in beyond Horndeski. 
Both solutions admit secondary hair and are asymptotically flat while the scalar field asymptotically decays.
For the black hole in the Horndeski class, we also give an extension of the solution to include time dependence.  

\subsection{Quartic Horndeski square root Lagrangian}
\label{sec:sqrt}

Following the method stemming from (\ref{eq:nanalytic}), we consider the Lagrangian (\ref{action}) with 
\bea
G_2 &=& \eta X,
\nn
G_4 &=& \zeta + \beta \sqrt{-X},
\nn
G_3 &=&  G_5= F_4=F_5=0,
\eea
where $\eta$ and $\beta$ are dimensionless parameters, and $\zeta=M_\tx{Pl}^2/(16 \pi)$. Note that $\eta$ or $\beta$ could be absorbed in a redefinition of the scalar field. We will not do so, in order to keep track of the origin of the various terms. 
The $G_2$ term is simply a canonical kinetic term, and the coefficient $\zeta$  in $G_4$ yields an Einstein-Hilbert piece in the action. 
The $\beta \sqrt{-X}$ term is in the class defined by~(\ref{eq:nanalytic}) and gives a $\phi$-independent contribution to the current. For completeness, let us write explicitly the action:
\be
S = \displaystyle\int{\mathrm{d}^4x \sqrt{-g} \left\{ \left[\zeta + \beta \sqrt{(\pd \phi)^2/2}\right] R - \dfrac{\eta}{2} (\pd \phi)^2 - \dfrac{\beta}{\sqrt{2(\pd \phi)^2}} \left[(\Box \phi)^2-(\nd_\mu \nd_\nu \phi)^2\right] \right\}}.
\label{eq:action}
\ee
It is interesting to note in passing that the above action for $\zeta=0$ admits global scale invariance, as was shown in~\cite{Padilla:2013jza}.

\subsubsection{Static solution}

Using the ansatz~(\ref{eq:ansatz}) for the metric and scalar field, we obtain for the radial component of the current, Eq. (\ref{eq:Jrexpr}):
\be
J^r= \dfrac{\beta  \sqrt{2 f}}{r^2} \text{sgn}(\phi')-\eta  \phi' f.
\label{eq:currentsqrtjohn}
\ee
The first term does not depend on $\phi'$; it depends on its sign, but as we will see below, all solutions keep a fixed sign in the static region of the black hole. 
Solving Eq.~(\ref{eq:currentsqrtjohn}), we get
\be
\phi'=\pm \dfrac{\sqrt{2} \beta}{\eta  r^2 \sqrt{f}}.
\label{eq:phiprime}
\ee
This expression for $\phi'$ is real for $f>0$, i.e. outside of the black hole horizon. Applying the sgn function to the $J^r=0$ equation, we find that $\beta$ and $\eta$ necessarily share the same sign. Two other equations remain to be solved, namely the $(tt)$ and $(rr)$ components of Einstein equations (the $(\theta \theta)$ and $(\phi \phi)$ equations follow on from the previous ones). 
They can be found in appendix \ref{sec:apEOM}. 

For our particular model, the $(tt)$ equation is particularly simple to solve once Eq. (\ref{eq:phiprime}) has been used. It is actually a first order differential equation on $f$. The $(rr)$ equation then imposes that $h$ is equal to $f$ (up to an overall constant that simply amounts to a redefinition of time). The solution takes the following form:
\be
f(r) = h(r) = 1 -\dfrac{\mu}{r}-\dfrac{\beta ^2}{2 \zeta  \eta  r^2},
\label{eq:fsol1}
\ee
where $\mu$ is a free integration constant. Additionally, the kinetic density $X$ reads:
\be
X(r) = -\dfrac{\beta^2}{\eta^2 r^4}
\label{eq:Xstat}
,\ee
from which we can compute the scalar field. Because of shift symmetry, the scalar field is determined up to some constant. We can use this freedom to impose that $\phi$ vanishes at spatial infinity. Then, the solution depends on the sign of the parameters $\eta$ and $\beta$:
\bea
\phi(r) &=& \pm 2 \sqrt{\dfrac{\zeta}{\eta}} \left\{\tx{Arctan} \left[\dfrac{\beta ^2+\zeta  \eta  \mu  r}{\beta  \sqrt{2 \zeta  \eta  r (r-\mu )-\beta ^2}}\right]- \tx{Arctan}\left(\dfrac{\mu}{\beta}  \sqrt{\dfrac{\zeta \eta}{2}}\right) \right\}
\nn
&\tx{if}&~~ \beta>0~~ \tx{and}~~ \eta>0,
\\
\phi(r) &=& \pm 2 \sqrt{\dfrac{\zeta}{-\eta}} \left\{\tx{Argth}\left[\dfrac{\beta ^2+\zeta  \eta  \mu  r}{\beta  \sqrt{\beta ^2-2 \zeta  \eta  r (r-\mu )}}\right]+\tx{Argth}\left(\dfrac{\mu}{\beta}  \sqrt{\dfrac{-\zeta  \eta}{2}}\right)\right\}
\nn
&\tx{if}&~~ \beta<0~~ \tx{and}~~ \eta<0.
\label{eq:phisol}
\eea

The above solution describes a black hole with mass $\mu/2$. Note that the non-trivial scalar field backreacts on the metric in an interesting way: the spacetime solution is of the RN form. 
This is possibly related to the remnant of global conformal invariance shared by the action (\ref{eq:action}), as the spacetime metric solution has zero Ricci scalar curvature (as does RN spacetime).
Positive $\eta$ formally corresponds to an imaginary charge of the RN metric. In this case, we have an event horizon for any value of $\mu$ including that of zero (unlike RN spacetime). 
On the other hand, when $\eta$ is negative, the scalar field manifests itself in an electric-like contribution where $\sqrt{-\beta^2/(2 \zeta \eta)}$ plays a role similar to that of electric charge for spacetime. 
This `electric charge' is not an integration constant; it depends entirely on the parameters of the theory, which are fixed. 
Any such black hole experiences the exact same correction to the Schwarzschild metric. 
Choosing a negative $\eta$ significantly affects the inner structure of the black hole, but the solution is not to be trusted beyond the event horizon, as can be seen from the fact that $\phi'$ becomes imaginary there. For negative $\eta$, there exists a lower bound on $\mu$:
\be
\mu_\tx{min} = \sqrt{\dfrac{2\beta^2}{-\zeta \eta}},
\ee
which, when saturated gives an extremal black hole. 
Whenever $\mu<\mu_\tx{min}$, the solution does not describe a black hole any more, but rather a naked singularity. 
In terms of stability, positive $\eta$ corresponds to the `correct' sign in the standard kinetic term. 
The stability, however, also depends on the quartic Galileon term. Therefore, one cannot conclude on the stability of the solutions only by the sign of $\eta$, see e.g.~\cite{Deffayet:2010qz,Babichev:2012re}. We will leave such an analysis for future work.

The solution presented above is asymptotically flat. 
It then fulfills all assumptions of the no-hair theorem but hypothesis 1, due to the presence of $\sqrt{-X}$ in $G_4$. The metric features a Newtonian fall-off at spatial infinity. At $r\to \infty$, the scalar field decays as:
\be
\phi(r) \mathop{=}_{r \rightarrow \infty} \pm \dfrac{\sqrt{2} \beta}{\eta r} + \mathcal{O}(r^{-2}).
\label{eq:farphi}
\ee
This solution does not have primary hair, as no integration constant other than $\mu$ appears in (\ref{eq:farphi}). No electric-type charge is present at asymptotic infinity, rather it depends on the fixed parameters of the Lagrangian. The black hole manifestly has secondary hair due to the non-trivial scalar-tensor mixing.

To conclude about faraway asymptotics, let us remark that a cosmological constant can be added to the initial action. The solution is modified in the same way as it is in GR, and acquires anti-de Sitter or de Sitter asymptotics. Explicitly, setting $G_2 = \eta X-2\Lambda$ and $G_4 =\zeta + \beta \sqrt{-X}$, one gets 
\be
f(r) =h(r)= 1-\dfrac{\mu}{r} -\dfrac{\beta ^2}{2 \zeta  \eta  r^2} -\dfrac{\Lambda}{3 \zeta} r^2,
\ee
and the scalar field can still be computed from Eq.~(\ref{eq:phiprime}).

Let us now examine the near-horizon asymptotics. As a direct consequence of Eq.~(\ref{eq:phiprime}), the derivative of the scalar field diverges at the horizon. This is however a coordinate-dependent statement, which ceases to be true using the tortoise coordinate, for instance; the singularity is absorbed in the coordinate transformation. On the other hand, it is easy to check from Eq. (\ref{eq:phisol}) that the scalar itself is finite at the horizon. Crucially, $X$ does not diverge either close to the horizon, Eq. (\ref{eq:Xstat}). 
Also, since the metric is identical to RN solution, it is clearly regular. 
Therefore, all physically meaningful quantities are well behaved when approaching the horizon.

Finally, we should mention the interior of the black hole. 
There, $f<0$ and Eq. (\ref{eq:phiprime}) would imply that $\phi'$ becomes imaginary. 
This feature is not specific to our solution: all known static solutions possess it~\cite{Rinaldi:2012vy, Minamitsuji:2013ura, Anabalon:2013oea}. 
The reason for such a behavior is that $r$ and $t$ exchange their role as space and time coordinates when crossing the horizon. 

\subsubsection{Extension to the time-dependent case}
\label{sec:dyn}


In this section, we focus on the time-dependent solutions of theory (\ref{eq:action}), and endeavor to establish a link between these solutions and their static counterpart. 
The scalar field acquires a linear time dependence with velocity $q$:
\be
\phi =qt+\psi(r)
\label{eq:tdep},
\ee
instead of the static scalar field ansatz~(\ref{eq:ansatz}).
The equations of motion can be fully solved, yielding:
\bea
h&=& \dfrac{2 \zeta^2 q^2}{C}\left(1-\dfrac{\mu}{r}-\dfrac{\eta}{2 \zeta r} \int{\tx{d}r \: r^2X}\right),
\label{eq:hdyn}
\\
f&=& \dfrac{C}{2 \zeta^2 q^2} h,
\\
C&=&4\zeta^2 X\left(1-\dfrac{\eta}{\beta}r^2\sqrt{-X}\right)
\label{eq:Xdyn}
,\eea
where $C$ is an integration constant. We choose to impose $f=h$, i.e. we set $C=2 \zeta^2 q^2$. Then, $h$ is explicitly determined as a function of $X$, and $X$ must be found as the root of a third order algebraic equation. Indeed, in terms of $\sqrt{-X}$, Eq. (\ref{eq:Xdyn}) reads
\be
\left(\sqrt{-X}\right)^2 \left(1-\dfrac{\eta}{\beta}r^2\sqrt{-X}\right)= -\dfrac{q^2}{2}.
\label{eq:rhoconstr}
\ee 
The right-hand side of Eq. (\ref{eq:rhoconstr}) is always negative. Therefore, solutions exist only when $\eta$ and $\beta$ share the same sign, similarly to the static case. When $\eta$ and $\beta$ have the same sign, it is easy to see that the above equation has a single positive root for all $r$. It reads:
\be
\sqrt{-X} = \dfrac{\beta}{3 \eta r^2}\left(1-A^{1/3}-A^{-1/3}\right),
\label{eq:rho}
\ee
where
\be
A = \dfrac12 \left[\sqrt{-4+\left(\dfrac{27 q^2 \eta^2 r^4}{2 \beta^2} +2\right)^2} -2 - \dfrac{27 q^2 \eta^2 r^4}{2 \beta^2}\right].
\ee
Let us point out that the $q=0$ solution nicely fits in these expressions. Indeed, in such a limiting case, $A=-1$ and $\sqrt{-X}=\beta/(\eta r^2)$, which is exactly what we found, Eq. (\ref{eq:Xstat}). The expression (\ref{eq:hdyn}) for $h$ also remains unchanged when $q=0$. 

The presence of a nonzero velocity completely changes however the asymptotics of the solution. Performing an expansion at large $r$ for a nonvanishing velocity $q$, we find that
\be
h \mathop{\sim}_{r \rightarrow \infty} \dfrac{3 \eta}{10 \zeta} \left(\dfrac{q^2 \beta}{2 \eta}\right)^{2/3} r^{2/3}
.\ee
Hence, if we drop the staticity assumption, we must abandon asymptotic flatness at the same time. The power 2/3 in $h$ was already found for a particular subclass of the `John' solutions in \cite{Babichev:2016rlq}. 
One should expect that when the velocity $q$ is small, one recovers the static solution. Let us keep $r$ fixed and expand $h$ in terms of $q$:
\be
h \mathop{=}_{q \rightarrow 0}1-\dfrac{\mu}{r}-\dfrac{\beta^2}{2 \eta \zeta r^2}+\dfrac{\eta q^2 r^2}{24 \zeta} + \mathcal{O}(q^4)
.\ee
The $q^2$ correction is negligible as long as
\be
r \ll r_q, ~~~~~~r_q \equiv \sqrt{\dfrac{\beta}{\eta q}}.
\ee
Physically, it means that the static solution approximates very well its time-dependent counterpart in the range $r_\tx{h} < r \ll r_q$ ($r_\tx{h}$ corresponding to the horizon of the static solution). Of course, this is true provided such a range exists, i.e. for small enough $q$. In this case, the presence of a nonzero velocity only affects the asymptotics of the solution; the details of the metric close to the black hole are unchanged.

The scalar field, on the other hand, is also modified, even at small radii. The introduction of time dependence actually makes the scalar field more regular close to the horizon. It stays real beyond the horizon, at least up to some non-vanishing depth. This can be seen thanks to the expression of $\phi'$, which we can get from $X$:
\be
X=\dfrac12\left(\dfrac{q^2}{h}-f \phi'^{\:2}\right)
.\ee
Then, using our solution, Eq. (\ref{eq:rho}), we extract $\phi'$:
\be
f^2 \phi'^{\:2} = q^2 + f \dfrac{2 \beta^2}{9 \eta^2 r^4} \left(1-A^{1/3}-A^{-1/3}\right)^2
.\ee
Like in the static case, $\phi'^{\:2}$ is clearly positive outside of a black hole. Furthermore, $A$ remains finite and never vanishes; hence, the $f$ part in the right-hand side tends towards zero when crossing the horizon. Thanks to the presence of $q^2$, the overall right-hand side remains positive inside the black hole, at least close to the horizon. 

\subsection{Quartic beyond Horndeski Lagrangian}
\label{sec:quarticbH}

A very similar analysis can be carried out for the beyond Horndeski quartic function $F_4$. Following Eq.~(\ref{eq:nanalytic}), we consider the following Lagrangian:
\bea
G_2 &=& \eta X,
\nn
G_4 &=& \zeta,
\nn
F_4 &=& \gamma(-X)^{-3/2},
\nn
G_3 &=& G_5=F_5=0,
\eea
where $\eta$ and $\zeta$ have the same meaning as in paragraph~\ref{sec:sqrt}, and $\gamma$ is a new dimensionful parameter. We follow the same steps as in the previous paragraph. The $J^r=0$ equation provides an expression for the kinetic density $X$:
\be
X=-\left[\dfrac{4 \gamma}{\eta} \; \dfrac{f(rh)'}{r^2 h}\right]^2
\label{eq:XbH}
.\ee
Then, using the $(rr)$ equation (\ref{eq:Err}), we can determine a particular combination of $f$ and $h$:
\be
\dfrac{f(rh)'}{r^2 h}=\dfrac{\eta}{48\gamma^2} \left[-\zeta+\sqrt{\zeta^2+\dfrac{96 \zeta \gamma^2}{\eta r^2}}\right]
\label{eq:combinationbH}
.\ee
Substituting this into the $(tt)$ equation, one ends up with a first order differential equation on $f$, the solution of which is
\be
f=\dfrac{1}{r(4\gamma\sqrt{-X}-\zeta)^2}\left[C-\displaystyle\int{\tx{d}r\left(\zeta+\dfrac12 \eta r^2 X\right)(4\gamma \sqrt{-X}-\zeta)}\right]
,\ee
with $C$ a free integration constant. $X$ is known in terms of $r$, by combining Eqs.~(\ref{eq:XbH}) and~(\ref{eq:combinationbH}). We can also compute $h$ from Eq. (\ref{eq:combinationbH}). The explicit expression for $f$, though not very enlightening, can be found in appendix~\ref{sec:apF4}. Again, this solution is asymptotically flat, with a Newtonian fall-off. Taking for instance a positive $\eta$, and defining the quantity
\be
\mu= \dfrac{-9C\sqrt{\eta}+4\sqrt{6} \gamma \zeta^{3/2} \:\tx{ln} (6\gamma^2 \zeta^3 \eta)}{9 \zeta^2 \sqrt{\eta}}
,\ee
we can expand $f$ at spatial infinity and get
\be
f(r) \mathop{=}_{r \rightarrow \infty} 1-\dfrac{\mu}{r}+\dfrac{40 \gamma^2}{\zeta  \eta  r^2} + \mathcal{O}(r^{-3})
.\ee
Therefore, $\mu$ should be interpreted as twice the mass of the black hole. The solution is very similar to what we obtained when considering $G_4\propto\sqrt{-X}$.

\section{Linear coupling to the Gauss-Bonnet invariant}
\label{sec:GB}
 
This section is devoted to a second model that a priori bifurcates the no-hair theorem, again following (\ref{eq:nanalytic}). The difference is that it involves the quintic sector of Horndeski theory rather than the quartic one. Namely, we choose the following Lagrangian:
\bea
G_2 &=& \eta X,
\nn
G_5 &=& \alpha \: \tx{ln}|X|,
\nn
G_4&=&\zeta,
\nn
G_3 &=&F_4=F_5=0,
\label{eq:GBlag}
\eea
where $\alpha$ is a new dimensionful coupling constant; it is well known that this particular $G_5$ Lagrangian can be written as $\mathcal{L}_\tx{GB} = -\alpha \phi \hat{G}/4$, with $\hat{G}$ the Gauss-Bonnet  invariant:
\be
\hat{G}=R_{\mu \nu \rho \sigma}R^{\mu \nu \rho \sigma} - 4 R_{\mu \nu} R^{\mu \nu} +R^2
.\ee
Reference~\cite{Sotiriou:2014pfa} recently studied the no-hair theorem and black holes in this theory. Black holes involving the Gauss-Bonnet invariant were found in the past (see for example~\cite{Kanti:1995vq} and~\cite{Campbell:1991kz}) due to effective string theory actions (but also no-hair theorems).
Here, we will follow a different approach to that of previous works because we will impose that the radial current vanishes: $J^r=0$.
In the previous works, although not stated,  $J^\mu$ does not vanish and as a result its norm is infinite for the black hole solutions discussed extensively in \cite{Sotiriou:2014pfa}. 
A priori, one would expect that $J^2$ is finite at the horizon, which is one of the key physical hypothesis of the no-hair theorem. We will see however that for this class of theories, $J^r=0$ does not lead to black hole solutions and one must have infinite current in order to find a static black hole with a static scalar. This either points to a pathology of this theory or to a particularity rendering the finiteness of the current irrelevant. 

First, we take a look at the spatial infinity expansion of the solution, assuming that it can be expanded in a $1/r$ series. We find that
\bea
h &\vphantom{=}& \mathop{=}_{r\rightarrow\infty} 1 -\dfrac{\mu}{r} - \dfrac{2\alpha ^2 \mu ^3}{7\zeta  \eta  r^7} + \mathcal{O}(r^{-8}),
\nn
f &\vphantom{=}& \mathop{=}_{r\rightarrow\infty} 1 -\dfrac{\mu}{r} - \dfrac{\alpha ^2 \mu ^3}{\zeta  \eta  r^7} + \mathcal{O}(r^{-8}),
\nn
\phi' &\vphantom{=}& \mathop{=}_{r\rightarrow\infty} -\dfrac{\alpha \mu^2}{\eta r^5} + \mathcal{O}(r^{-6}),
\label{eq:farGB}
\eea
with $\mu$ a free integration constant. The corrections with respect to GR are therefore very mild far away from the source. These corrections are in agreement with the post-Newtonian corrections for a distributional source found in \cite{Amendola:2007ni}. In contrast with \cite{Sotiriou:2014pfa}, we notice that $\phi$ decays as $1/r^4$ and the only free parameter is the mass of the central object, $\mu$. There is no tunable scalar charge as expected, since $J^r=0$ is already an integral of the scalar equation of motion. The above expansion cannot be trusted whenever the $\alpha^2$ corrections become of the same order as the mass term, i.e. when
\be
r \lesssim \left(\dfrac{\alpha^2 \mu^2}{\zeta \eta}\right)^{1/6}
.\ee

\begin{figure}[ht]
\begin{center}
\includegraphics[width=10cm]{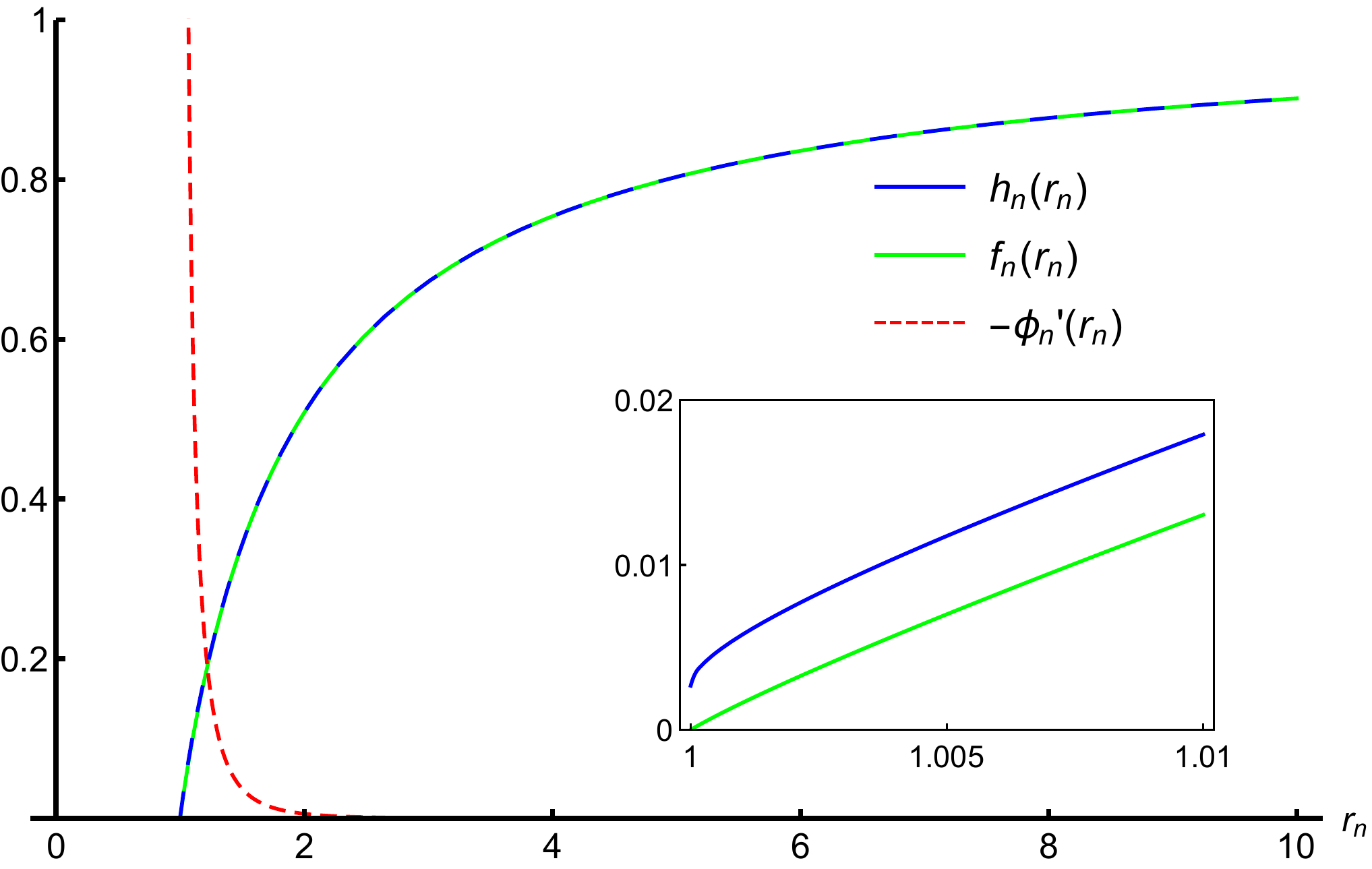}
\caption{Typical numerical solution. All quantities indexed by n refer to the corresponding quantities in the system of dimensionless units used for the integration, see appendix~\ref{sec:apGB}. $f_\tx{n}$ and $h_\tx{n}$ are in very good agreement with the spatial infinity expansion at large values of $r_\tx{n}$. However, the zoomed plot reveals a pathological behavior close to the point where $f_\tx{n}$ vanishes. It is clear that $h_\tx{n}$ does not vanish at the same time.}
\label{fig:coordsing}
\end{center}
\end{figure}

To go further, we resort to numerical integration, because the system of equations was not integrated analytically. The details of this integration can be found in appendix~\ref{sec:apGB}. Here, we simply summarize the important results of this analysis. In brief, the procedure consists in imposing that $f$ vanishes at some given radius, and integrating outwards. A typical result of the numerical integration is displayed in Fig.~\ref{fig:coordsing}.

Far away, the metric and scalar field fit very well the expansion given in Eq.~(\ref{eq:farGB}). However, taking a closer look at the black hole region itself, one remarks that $h$ does not vanish when $f$ does, as should be the case for a black hole. This behavior is shown in Fig.~\ref{fig:coordsing}, and we also confirmed this by an analytical expansion close to the point where $f$ vanishes. All curvature invariants being finite there, this suggests the presence of a coordinate singularity. To go through this coordinate singularity, we implemented a change of coordinates, detailed in appendix~\ref{sec:apGB}. This indeed allows us to go beyond the coordinate singularity; however, close enough, the numerical simulation breaks down while the Ricci scalar blows up at the same time. Therefore, the solution describes a naked singularity, no horizon being present.

We conclude that there are no black hole solutions for the Lagrangian (\ref{eq:GBlag}), if one requires that the norm of the current is finite. Far away from the curvature singularity though, the metric can describe the exterior of a star. A similar behavior was found in~\cite{Charmousis:2015txa} in a different set-up.

\section{Purely quartic models}
\label{sec:quartic}

In this section, we explore an alternative way to construct hairy black hole solutions. 
We wish to respect hypothesis~1 in~(\ref{eq:hyp}), but to break hypothesis~2; that is, instead of considering non-analytic $G_i$ functions as in previous sections, we keep regular functions but at the same time we set 
$$
G_2=G_3=G_5=F_5=0,
$$
with arbitrary $G_4$ and $F_4$. Doing so, we automatically get rid of the canonical kinetic term, so that the no-hair theorem does not apply any more. 
The scalar field and the metric are now assumed to be static, Eq. (\ref{eq:ansatz}). 
The equations of motion involve the density $X$ only (see Appendix~\ref{sec:apEOM}). 
Extracting the combination $(rh)'f/h$ from both the $(rr)$ equation and the $J^r=0$ equation and equating the two expressions, 
we are left with
\be
\dfrac{G_{4X}}{G_{4X}+2XG_{4XX}+42X^2F_{4X}+8XF_4} =\dfrac{G_4}{G_4-2XG_{4X}-4X^2F_4} 
\label{eq:Xconst}
.\ee
It is remarkable that the above equation does not involve the radial parameter $r$. 
Equation~(\ref{eq:Xconst}) should be understood as an equation on $X$ for a fixed choice of  $G_4$ and $F_4$. 
Let us assume for now that Eq.~(\ref{eq:Xconst}) has a solution, $X=X_0$. 
The fact that~(\ref{eq:Xconst}) does not involve $r$ means that $X$ is constant everywhere. 
This greatly simplifies the $(tt)$ equation, which can be immediately integrated. The solution reads
\bea
h(r)&=& 1-\dfrac{\mu}{r},
\nn
f(r) &=& \left(1-\dfrac{\mu}{r}\right) \dfrac{G_4(X_0)}{G_4(X_0)-2 X_0 G_{4X}(X_0)-4X_0^2F_4(X_0)},
\eea
where $\mu$ is an integration constant and keeping in mind that $X_0$ must be a solution of Eq.~(\ref{eq:Xconst}). Therefore, the static and spherically symmetric solutions of a fully general quartic Horndeski theory boil down to a simple Schwarzschild metric, up to a solid angle deficit (corresponding to the constant in front of $f$). 
We can avoid a solid deficit angle (which would lead to a curvature singularity even for $\mu=0$!) by requiring an extra condition on the functions $G_4$ and $F_4$, such that the extra factor in front of $f$ is 1. 
The combination of this condition with Eq.~(\ref{eq:Xconst}) gives
\bea
0&=&G_{4X}(X_0)+2X_0F_4(X_0),
\nn
0&=&G_{4XX}(X_0)+4F_4(X_0)+2X_0F_{4X}(X_0),
\label{eq:constr}
\eea
for some value $X=X_0$. 

We have thus found infinitely many theories that possess a stealth Schwarzschild black hole solution. Namely, all those which fulfill the constraints given in Eq. (\ref{eq:constr}) at some point $X_0$. A particularly interesting class among these is the subspace of $\{G_4,\:F_4\}$ theories where $F_4=0$. In this subclass, the models that possess such a stealth black hole are the theories with $G_{4X}(X_0)=0$ and $G_{4XX}(X_0)=0$. Any theory of the type
\be
G_4(X)=\zeta+\sum_{n\geq2} \beta_n (X-X_0)^n
\label{example:stealth}
\ee
will allow for a Schwarzschild metric with a non-trivial scalar field. A more general examination of theories having $X=X_0$ with $G_{4X}(X_0)=0$ and $G_{4XX}(X_0)=0$ shows that any such theory allows for all Ricci-flat solutions, with a non-vanishing hidden scalar field. For instance, these theories admit as a solution the Kerr metric (here in Boyer-Lindquist coordinates):
\bea
\tx{d} s^2 = &-& \left(1-\dfrac{2 m r}{r^2+a^2 \cos^2 \theta}\right)\tx{d} t^2 -\dfrac{4 m r a \sin^2 \theta}{r^2+a^2 \cos^2 \theta} \tx{d}t \tx{d}\phi +\dfrac{r^2+a^2 \cos^2\theta}{r^2-2 m r +a^2}\tx{d} r^2
\nn
&+& \tx(r^2+a^2 \cos^2\theta){d}\theta^2 + \left(r^2+a^2+\dfrac{2 m r a^2 \sin^2\theta}{r^2+a^2 \cos^2\theta}\right)\sin^2 \theta \tx{d}\varphi^2
,\eea
with a scalar field given by
\be
\phi (r,\theta)= \sqrt{-2 X_0} \left[ a\sin \theta- \sqrt{a^2-2 m r+r^2}-m~\tx{ln} \left(\sqrt{a^2-2 m r+r^2}-m+r\right)\right]
,\ee
$a$ being the rotation parameter and $m$ the mass of the black hole. This scalar field is regular everywhere outside of the event horizon of the Kerr black hole.

A remarkable characteristic of this class of solutions is that, even though the geometry is asymptotically flat, the scalar does not vanish at spatial infinity: 
its derivative $\phi'$ tends towards a finite constant.
This violates another assumption of the no-hair theorem: it is required that $\phi'\rightarrow 0$ at spatial infinity. Therefore, the class of solutions we are dealing with breaks two hypotheses. 

The black hole solutions found in this section are reminiscent of the properties of the ghost condensate in the field of a black hole~\cite{Mukohyama:2005rw}. 
Indeed, for this theory, which contains only a non-trivial function~$G_2(X)$ (while other functions are zero) with a minimum at some $X=X_0$, 
the situation is very similar. At the point $X=X_0$ the energy momentum tensor for this theory becomes equivalent to that of the cosmological term. 
Adjusting $G_2(X)$ in such a way that the cosmological term is zero, one gets a stealth black hole solution, similar to our solutions in this section. 
In the case of the $G_2(X)$ theory, there is a pathology though --- the theory becomes non-dynamical at the point $X=X_0$. 
A way to overcome this pathology is to introduce higher-order terms.
Therefore it is still to be understood, whether a theory~(\ref{eq:constr}) is healthy at the point $X=X_0$. 
We leave this study for future work.

\section{Conclusions}
\label{sec:conc}

We studied extensively black holes with secondary hair in shift-symmetric Horndeski and beyond Horndeski theories. We have concentrated mainly on finding static and spherically symmetric solutions with a static scalar field. In certain cases we investigated how solutions are extended if we allow for a mild linear time dependence in the scalar field. The central property in the theories under study is
shift symmetry which generates a conserved Noether current, the key quantity for our analysis. 
Indeed, making the physically reasonable assumption that the norm of the current is finite~\cite{Hui:2012qt}, 
along with asymptotic flatness and two technical requirements on the form of the Lagrangian, a no-hair theorem can be proven~\cite{Hui:2012qt}.
The starting point in our research were Lagrangians, which on one hand have a current of finite norm, but on the other do not respect one of the latter two technical requirements. We thus identified possible candidate theories with hairy black holes.

A first class among those Lagrangians is built using specific functions $G_i$ and $F_i$ as in (\ref{eq:nanalytic}). 
In fact, $\phi'$ has to be non-trivial in order to achieve a finite norm of the current.
We found six different models in the beyond Horndeski theory (four if restricted to the Horndeski theory) which possibly admit hairy black holes, see Eq.~(\ref{eq:nanalytic}).
As two illustrative examples, we found black hole solutions with secondary hair in a subclass of quartic Horndeski and beyond Horndeski theory, Sec.~\ref{sec:nanalytic}. 
Within shift-symmetric (beyond) Horndeski theories, this is the first exact black hole solution that is static, asymptotically flat and that has finite norm of the current. 
All observable quantities made of the metric and the scalar field are well-behaved, both at spatial infinity and at the horizon. 
The scalar field decays like $1/r$, and backreacts on the metric through a rapidly damped contribution in $1/r^2$. 
These models provide an interesting link between the static and time-dependent solutions built in the way proposed in~\cite{Babichev:2013cya,Kobayashi:2014eva}. 
Indeed we extended one class of solutions to include a linearly time-dependent scalar field (the only extension allowed for the specific spacetime symmetry) 
and we saw that the extended solutions change asymptotic behavior. 

Within our study, we also investigated in detail the theory with a linear coupling between the scalar field and the Gauss-Bonnet density. 
This case presents some interest due to the fact that the Gauss-Bonnet term is a topological invariant in the absence of a scalar field. 
Black holes in this theory have been studied recently~\cite{Sotiriou:2014pfa}, in connection 
to the no-hair theorem of Ref.~\cite{Hui:2012qt} but also in the past (see for example \cite{Kanti:1995vq} and \cite{Campbell:1991kz}).
Interestingly, this Lagrangian is in the family under study (\ref{eq:nanalytic}) whereupon the scalar field is sourced by the Gauss-Bonnet curvature scalar away from the trivial configuration. The black hole solutions
however, have a singular norm for the conserved current at the horizon \cite{Babichev:2016rlq}. In Sec.~\ref{sec:GB}, we looked for solutions in this theory 
that have a regular norm for the current. Within our study, no solutions with a regular event horizon were found, although the solutions asymptotically agree with Dirac sourced star solutions found previously in this theory \cite{Amendola:2007ni}. This means that either the finiteness of the norm is not relevant (for black hole solutions), or points at some pathology of this theory. This demands further study.

A second way to build black holes with hair, presented in Sec.~\ref{sec:quartic}, is to remove the canonical kinetic term from the action. Indeed, in this way a stealth Schwaschild solution was found in the presence of a linearly time-dependent scalar field \cite{Babichev:2013cya}. Time dependence for the scalar was essential there in order to have a non trivial (regular) scalar field. Here we allowed no time dependence but left arbitrary $G_4$ and $F_4$ in quartic (beyond) Horndeski theory.  
We obtained the generic solution, which is described by a Schwarzschild metric with a solid deficit angle. 
The scalar field is not trivial, and does not vanish at spatial infinity. 
We found simple conditions in order to have a regular black hole. In the case of $G_4$ alone, inspection of the field equations shows that the two conditions imposed on the theory actually allow for any Ricci-flat solution with a constant density $X$. The stationary Kerr metric is indeed solution to the field equations, with a non-trivial scalar field profile.

We left aside the study of perturbations around the solutions we presented. 
Such a study should reveal possible hidden pathologies of the theories or/and solutions. 
Indeed, an unpleasant feature of the family of Lagrangians~(\ref{eq:nanalytic}) is that they are non-analytic around Minkowski vacuum. 
This peculiarity, unseen at the level of solutions, may become present at the level of perturbations. 
We leave this and other related questions for future work.

%

 \acknowledgments
We are grateful to Marco Crisostomi and Gilles Esposito-Far\`ese for interesting discussions. The authors acknowledge financial support from the research program, Programme
national de cosmologie et galaxies of the CNRS/INSU, France. EB was supported in part by Russian Foundation for Basic Research Grant No. RFBR 15-02-05038.

\appendix
\section{Equations of motion for a spherically symmetric and static ansatz}
\label{sec:apEOM}

For our purposes, it is necessary and sufficient to make use of two of the equations of motion, the $(tt)$ and $(rr)$ ones. We assume that both the metric and the scalar field are spherically symmetric and static. We remove the cubic part of Horndeski and quintic part of beyond Horndeski actions, since these are cumbersome expressions that we did not use in the course of the paper. The full expressions, including $G_3$ and $F_5$, can be found in \cite{Babichev:2016kdt}, at the cost of a tedious notation translation. Then, the $(tt)$ equation reads:
\bea
\label{eq:Ett}
&G_2& + \dfrac{2}{r} \left(\dfrac{1 - f}{r} - f'\right) G_4 + 4 \dfrac{f}{r} \left(\dfrac{1}{r} + \dfrac{X'}{X} + \dfrac{f'}{f}\right) X G_{4X} + 8 \dfrac{f}{r} X X' G_{4XX} 
\nn
&+& \dfrac{f\phi'}{r^2}\left[(1-3f)\dfrac{X'}{X}-2f'\right]XG_{5X}-\dfrac{2}{r^2}f^2\phi'XX'G_{5XX}+\dfrac{16f}{r}X^2X'F_{4X}
\nn
&+&\dfrac{8f}{r}\left(\dfrac{4X'}{X}+\dfrac{f'}{f}+\dfrac{1}{r}\right)X^2F_4 =0,
\eea
To be precise, rather  than the $(rr)$ equation itself, we use a linear combination of it with the $J^r=0$ equation, in the fashion described in \cite{Babichev:2016kdt}. The result gives:
\bea
\label{eq:Err}
&G_2& - \dfrac{2}{r^2 h} (frh'+fh-h) G_4 + \dfrac{4f}{r^2h} (rh'+h) X G_{4X} - \dfrac{2}{r^2h}f^2h'\phi'XG_{5X}
\nn
&+&\dfrac{8f}{r^2h}(rh'+h)X^2F_4 =0.
\eea

\section{Metric function for the quartic beyond Horndeski Lagrangian solution}
\label{sec:apF4}

The explicit expression of the metric function $f$ for the solution described in paragraph \ref{sec:quarticbH} reads:
\bea
 f(r)= &\vphantom{\left[ \dfrac{r}{\gamma \zeta  \left(24 \gamma+\sqrt{6} \sqrt{96 \gamma^2+\zeta  \eta  r^2}\right)}\right]}& \dfrac{1}{144 \gamma ^2 \eta  r \left\{2 \zeta +\sqrt{\zeta [\zeta + 96 \gamma ^2/(\eta r^2)]}\right\}^2 \left(96 \gamma ^2+\zeta  \eta  r^2\right)} \left\{48 \gamma^2 r^2 \zeta \eta \vphantom{\left[ \dfrac{r}{\gamma \zeta  \left(24 \gamma+\sqrt{6} \sqrt{96 \gamma^2+\zeta  \eta  r^2}\right)}\right]} \right.
\nn
&\vphantom{\left[ \dfrac{r}{\gamma \zeta  \left(24 \gamma+\sqrt{6} \sqrt{96 \gamma^2+\zeta  \eta  r^2}\right)}\right]}& \times \left[10\zeta \sqrt{\zeta \eta (96 \gamma^2+\zeta  \eta  r^2)}+\eta (27 C+16 \zeta ^2 r)\right]+r^4 \zeta^3 \eta^2\left[\sqrt{\zeta \eta (96 \gamma^2+\zeta  \eta  r^2)}-\zeta \eta r\right]
\nn
&\vphantom{\left[ \dfrac{r}{\gamma \zeta  \left(24 \gamma+\sqrt{6} \sqrt{96 \gamma^2+\zeta  \eta  r^2}\right)}\right]}& +\:4608 \gamma^4 \left[8 \zeta \sqrt{\zeta \eta (96 \gamma^2+\zeta  \eta  r^2)}+27 C \eta +18 \zeta ^2 \eta  r\right]+1152 \sqrt{6} \gamma^3 \zeta^{3/2} \sqrt{\eta } (96 \gamma^2+\zeta  \eta  r^2) 
\nn
&\vphantom{\left[ \dfrac{r}{\gamma \zeta  \left(24 \gamma+\sqrt{6} \sqrt{96 \gamma^2+\zeta  \eta  r^2}\right)}\right]}& \left. \times \: \tx{ln} \left[ \dfrac{r}{\gamma \zeta  \left(24 \gamma+\sqrt{6} \sqrt{96 \gamma^2+\zeta  \eta  r^2}\right)}\right]\right\}
,\eea
where we assumed that $\eta$ was positive ($\eta$ and $\gamma$ must have opposite sign).

\section{Numerical analysis of the linear coupling to Gauss-Bonnet density}
\label{sec:apGB}

In this appendix, we detail the numerical analysis we implemented to study the solutions of the theory~(\ref{eq:GBlag}).
For this particular theory, the radial component of the current reads
\be
J^r= f \left[\dfrac{\alpha  (f-1) h'}{r^2 h}-\eta \phi' \right],
\label{eq:GBcurrent}
\ee
and we still use the $(tt)$ and $(rr)$ equations of appendix \ref{sec:apEOM}. First, imposing $J^r=0$, we can extract $h'/h$ as a function of $\phi'$ and $f$:
\be
\dfrac{h'}{h}=\dfrac{\eta r^2 \phi'}{\alpha(f-1)}
\label{eq:hGB}
.\ee
Using Eq. (\ref{eq:hGB}), the $(rr)$ equation becomes a second-order algebraic equation on $\phi'$; we solve it and get $\phi'$ in terms of $f$:
\be
\phi'=\dfrac{-2 \zeta  \eta  r^3 f \pm\sqrt{4 \zeta ^2 \eta ^2 r^6 f^2- 4 \alpha^2 \eta \zeta r^2 f (1 -f)^2(5 f-  1)}}{\alpha \eta r^2 f (5 f-1)}
\label{eq:phiGB}
.\ee
Two branches exist for $\phi'$. Relying on the numerical analysis we performed, we select the `plus' branch; the `minus' branch gives pathological solutions that extend only to a finite radius. Equation~(\ref{eq:phiGB}) also fixes the sign of $\eta$. Indeed, taking the limit $f \rightarrow 0$, as expected for a black hole, one can check that the sign of the term under the square root is determined by the sign of $\eta$ at leading order ($\zeta$ is positive by convention). If $\eta$ was negative, the scalar field would become imaginary before reaching the assumed horizon. Therefore, we will restrict the analysis to positive $\eta$.

We are left with one master equation on $f$, which turns out to be a first-order ordinary differential equation. 
To write it in a form adapted to numerical resolution, we introduce a length scale $r_0$, and consider functions of $r_\tx{n} \equiv r/r_0$, rather than $r$. 
Then, the master equation depends merely on one dimensionless parameter, that we call $\alpha_\tx{n}$:
\be
\alpha_\tx{n} \equiv \dfrac{\alpha}{\sqrt{\eta \zeta} r_0^2}
.\ee
We use the following dictionary between dimensionless and dimensionful quantities:
\bea
f_\tx{n} (r_\tx{n}) = f(r_\tx{n} r_0),\quad  h_\tx{n} (r_\tx{n}) = h(r_\tx{n} r_0),\quad \phi_\tx{n} (r_\tx{n}) = \sqrt{\dfrac{\eta}{\zeta}} \phi(r_\tx{n} r_0).
\eea
The master equation in terms of $f_\tx{n}$ reads:
\bea
\label{eq:masterf}
&\vphantom{\dfrac12}&4 f_\tx{n}^3 \left(40 r_\tx{n} \alpha_\tx{n} ^4 f_\tx{n}'+5 r_\tx{n}^4 \alpha_\tx{n} ^2+53 \alpha_\tx{n} ^4\right)+r_\tx{n} \alpha_\tx{n} ^2 \left(-12 r_\tx{n} S + r_\tx{n}^4+\alpha_\tx{n} ^2\right) f_\tx{n}'
\nn
&\vphantom{\dfrac12}&-2 f_\tx{n} \left[-r_\tx{n} \alpha_\tx{n} ^2 \left(10 r_\tx{n} S+r_\tx{n}^4+4 \alpha_\tx{n} ^2\right) f_\tx{n}'+6 r_\tx{n} \alpha_\tx{n} ^2 S+r_\tx{n}^8-8 r_\tx{n}^4 \alpha_\tx{n} ^2-13 \alpha_\tx{n} ^4\right]
\nn
&\vphantom{\dfrac12}&+\alpha_\tx{n} ^2 f_\tx{n}^2 \left[\left(5 r_\tx{n}^5-94 r_\tx{n} \alpha_\tx{n} ^2\right) f_\tx{n}'+10 r_\tx{n} S-34 r_\tx{n}^4-116 \alpha_\tx{n} ^2\right]
\nn
&\vphantom{\dfrac12}&-5 \alpha_\tx{n} ^4 f_\tx{n}^4 \left(15 r_\tx{n} f_\tx{n}' +34\right)+2 \left(r_\tx{n}^4+\alpha_\tx{n} ^2\right) \left(r_\tx{n} S-\alpha_\tx{n} ^2\right)+50 \alpha_\tx{n} ^4 f_\tx{n}^5 = 0,
\eea
where
\be
S=\sqrt{r_\tx{n}^2 f_\tx{n} \left\{f_\tx{n} \left[\alpha_\tx{n} ^2 f_\tx{n} (11-5 f_\tx{n})+r_\tx{n}^4-7 \alpha_\tx{n} ^2\right]+\alpha_\tx{n} ^2\right\}}.
\ee
Since this is a first-order differential equation, we need to specify one initial condition. Because we are a priori looking for a black hole, we impose that $f_\tx{n}$ vanishes at $r_\tx{n}=1$. Then we can proceed to numerical integration, and we obtain results similar to those of Fig.~\ref{fig:coordsing}. This particular plot was obtained for $\alpha_\tx{n}=1/10$. As mentioned in Sec.~\ref{sec:GB}, $h_\tx{n}$ does not vanish when $f_\tx{n}$ does, indicating a coordinate singularity. Such a pathological behavior was already observed in a similar theory by Kanti \textit{et al.} in \cite{Kanti:1995vq}. The theory they studied is a string-inspired exponential coupling of the scalar field to the Gauss-Bonnet invariant, namely the Lagrangian,
\be
R-\dfrac12 (\pd \phi)^2+\alpha \tx{e}^{-\gamma \phi} \hat{G}
\label{eq:dilatonlag}
.\ee
This theory is not shift-symmetric, but the characteristic features of the solution are similar to that derived from the theory~(\ref{eq:GBlag}).

\begin{figure}[ht]
\begin{center}
\includegraphics[width=10cm]{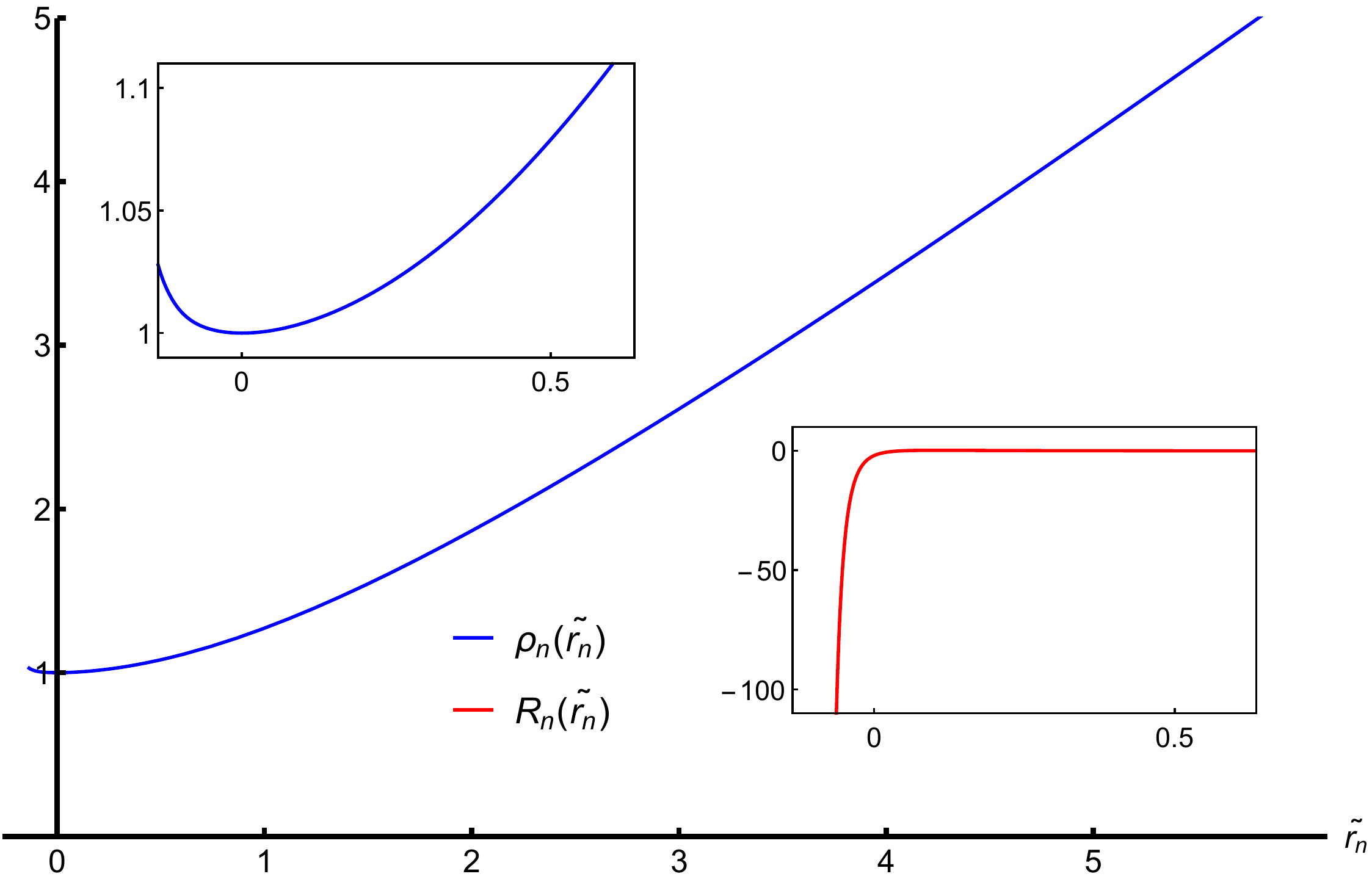}
\caption{Numerical solution obtained for $\alpha_\tx{n}=1/10$. At large $\tr_\tx{n}$, $\rho_\tx{n}(\tr_\tx{n}) \simeq \tr_\tx{n}$. At $\tr_\tx{n}=0$, the zoomed plot shows that $\rho_\tx{n}$ starts increasing again. However, the simulation breaks down at $\tr_\tx{n} \simeq -0.13$. The second framed plot shows that the Ricci scalar $R_\tx{n}$, in red, diverges at this point, indicating a curvature singularity.}
\label{fig:nakedsing}
\end{center}
\end{figure}

To go further, we must remark that there is no way to extend the solution in the region $r_\tx{n}<1$ because $\phi_\tx{n}'$ becomes imaginary there. Therefore we have to change coordinates. As the coordinate singularity arises from the $g_{rr}$ part of the metric, we define a new radial coordinate $\tr$ as
\be
\tx{d}\tr = \dfrac{\tx{d}r}{\sqrt{f(r)}}
.\ee
The ansatz for the metric now takes the following form:
\be
\tx{d}s^2 = -h(\tr) \tx{d}t^2 + \tx{d}\tr^2+\rho(\tr)^2\tx{d}\Omega^2
\label{eq:ansatz2}
,\ee
where $\rho$ is a new unknown function, interpreted as the areal radius, i.e. the radius that measures the area of constant $\tr$ 2-spheres. Repeating the same procedure as above, we eliminate $h$ and $\phi'$ in the equations and obtain a master equation on $\rho$ only, which is second order\footnote{The higher order of the equation translates the fact that the new ansatz~(\ref{eq:ansatz2}) involves an additional reparametrization freedom $\tr \rightarrow \tr + constant$.}. With the same convention as before for the units and the index n, this equation reads:
\bea
&\vphantom{\dfrac12}&50 \alpha_\tx{n} ^4 \rho_\tx{n}'^{\:11}-170 \alpha_\tx{n} ^4 \rho_\tx{n}'^{\:9}+\rho_\tx{n}^2 \tilde{S}^3-2 \alpha_\tx{n} ^2 \left(\rho_\tx{n}^4+\alpha_\tx{n} ^2\right) \rho_\tx{n}'+4 \left(5 \alpha_\tx{n} ^2 \rho_\tx{n}^4+53 \alpha_\tx{n} ^4\right) \rho_\tx{n}'^{\:7}
\nn
&\vphantom{\dfrac12}&-2 \left(17 \alpha_\tx{n} ^2 \rho_\tx{n}^4+58 \alpha_\tx{n} ^4\right) \rho_\tx{n}'^{\:5}+2 \left(8 \alpha_\tx{n} ^2 \rho_\tx{n}^4-\rho_\tx{n}^8+13 \alpha_\tx{n} ^4\right) \rho_\tx{n}'^{\:3}+2 \alpha_\tx{n} ^2 \rho_\tx{n} \rho_\tx{n}' \rho_\tx{n}'' \left[\rho_\tx{n}^4 \left(5 \rho_\tx{n}'^{\:4}+2 \rho_\tx{n}'^{\:2}+1\right)\vphantom{\left(\rho_\tx{n}'^2-1\right)^2}\right.
\nn
&\vphantom{\dfrac12}&\left.-\alpha_\tx{n} ^2 \left(\rho_\tx{n}'^{\:2}-1\right)^2 \left(75 \rho_\tx{n}'^{\:4}-10 \rho_\tx{n}'^{\:2}-1\right)\right]+\rho_\tx{n}^2 \tilde{S} \left\{\rho_\tx{n}'^{\:2} \left[\alpha_\tx{n} ^2 \left(15 \rho_\tx{n}'^{\:4}-23 \rho_\tx{n}'^{\:2}+9\right)\right. \right.
\nn
&\vphantom{\dfrac12}&\left. \left.+8 \alpha_\tx{n} ^2 \rho_\tx{n} \left(5 \rho_\tx{n}'^{\:2}-3\right) \rho_\tx{n}''+\rho_\tx{n}^4\right]-\alpha_\tx{n} ^2\right\}=0
,\eea
where
\be
\tilde{S}=\sqrt{\rho_\tx{n}'^{\:2} \left(\alpha_\tx{n} ^2 \left(-5 \rho_\tx{n}'^{\:4}+11 \rho_\tx{n}'^{\:2}-7\right)+\rho_\tx{n}^4\right)+\alpha_\tx{n} ^2}
.\ee
To proceed with numerical integration we specify two initial conditions. The configuration we impose is equivalent to the one of the previous paragraph: we require that $\rho_\tx{n}(0)=1$ and $\rho_\tx{n}'(0)=0$. In terms of the old ansatz, this would translate as $f_\tx{n}(1)=0$, and $r_\tx{n}=1$ is mapped to $\tr_\tx{n}=0$. The result is shown in Fig.~\ref{fig:nakedsing}.

We see that the solution can be continued in the $\tr_\tx{n}<0$ range, i.e. beyond the point where the old coordinate system becomes singular. In the new coordinates, the areal radius of 2-spheres $\rho$ decreases with $\tr$ up to $\tr=0$, and then starts increasing again when $\tr$ goes to negative values. However, the solution can be continued only in a very short range of negative $\tr$. When the integration fails, the Ricci scalar explodes; thus, our solution describes a curvature singularity which is not shielded by any horizon.

A second possibility though, is to interpret this type of solution as a wormhole. Indeed, for a similar solution, although for the different theory~(\ref{eq:dilatonlag}), this interpretation has been suggested in~\cite{Kanti:2011jz}. There are two branches of the solution for $\phi'$, and the idea of~\cite{Kanti:2011jz} amounts to gluing these two branches at $\tr=0$, so that the solution on the left is 
symmetrical to the solution on the right: $\phi(-\tr)=\phi(\tr)$, $\phi'(-\tr)=-\phi'(\tr)$, etc. One obtains to copies of the same asymptotically flat universe, glued together at $\tr=0$, where the areal radius $\rho$ is minimal. Doing this, one creates a throat that relates two universes, i.e. a wormhole. The price to pay for this is that certain quantities, namely $\phi'$ and $h'$, become discontinuous at $\tr=0$. This can be fixed, as proposed in~\cite{Kanti:2011jz}, by adding some matter located on the throat. If one tunes this matter to the right density and pressure, one can account for the discontinuities at the throat. 

\bibliographystyle{unsrt}
\bibliography{biblio}
\end{document}